\documentclass[aps,prd,preprint,nofootinbib]{revtex4-1}

\usepackage{amsmath,amssymb,slashed}
\usepackage{dsfont}
\usepackage{graphicx}
\usepackage{epstopdf}  
\usepackage{hyperref}

\begin{document}

\title{Spacelike zero crossings and timelike phases in the electromagnetic form factors of vector mesons}

\author{Stefan Leupold}

\affiliation{Institutionen f\"or fysik och astronomi, Uppsala universitet, Box 516, S-75120 Uppsala, Sweden}
\date{\today}

\begin{abstract}
  Some form factors of ground-state hadrons with spin might show a zero crossing at spacelike momenta of electron-hadron 
  scattering. In the timelike region of hadron-antihadron production by electron-positron collisions the form factors 
  become complex. When the hadrons decay, the relative phases between form factors can be measured by the angular distributions 
  of their decay products. Using analyticity and the QCD high-energy limits for constraint-free form factors and 
  for helicity amplitudes, we show how a single 
  zero crossing in the spacelike region is related to the high-energy limit of the phase in the timelike region. Theoretical 
  predictions for such zero crossings can therefore be tested by experimental measurements, e.g.\ by Belle II, 
  in the high-energy timelike region.   In the present work, this line of reasoning is applied to vector mesons. 
\end{abstract}

\maketitle

\section{Motivation}
\label{sec:motiv}

When the electric form factor of the proton was determined at Jefferson Lab, 
the data suggested that there might be a zero crossing at 
higher energies; see e.g.\ the discussion in the review \cite{Perdrisat:2006hj}. 
If true, one might qualitatively conclude that this form factor does not look ``as simple as possible'' (just falling off 
with energy), even though the proton is the ground state. It seems instead that the form factor falls fast, 
crosses zero, then turns around and approaches zero from below. 
As long as one has not confirmed this zero crossing experimentally, it might be interesting to consult also 
dispersion theoretical calculations (see \cite{Lin:2021umk} and references therein), which are based on
\begin{eqnarray}
  \label{eq:disp1}
  F(q^2) = \int\limits_0^\infty \frac{ds}{\pi} \, \frac{{\rm Im}F(s)}{s-q^2 - i \epsilon} \,.
\end{eqnarray}
We will not make detailed use of \eqref{eq:disp1} in the present work, but we have presented this relation to highlight that 
form factors have analytic properties, which in turn are rooted in the locality and causality principles of relativistic quantum 
field theory. This general aspect will be crucial to relate spacelike and timelike energy regions.

But suppose that there is a zero crossing in the electric form factor of the proton. 
Then one might want to investigate if there are also such zero crossings 
for other hadrons. How is it with the $\Lambda$ baryon or with vector mesons like $\rho$, $K^*$, $D^*$ or $B^*$? Do they 
have form factors that show a zero crossing? Does it depend on the flavor composition? 
For instance, there are quark-gluon Dyson-Schwinger calculations \cite{Xu:2019ilh} that relate meson and baryon form factors 
and suggest zero crossings in the respective electric form factors of vector mesons. 

For none of these other states $\Lambda$, $\rho$, $K^*$, \ldots , one can easily go to the spacelike momenta of electron-hadron 
scattering.\footnote{For the relatively long living $\Lambda$ one can imagine a $\Lambda$ beam directed on the electron cloud 
of atoms. But one cannot have $\rho$ or $K^*$ mesons as beams or targets.} 
But one can study other reactions, namely for timelike momenta: $e^+ e^- \to \Lambda \bar\Lambda$ or $\to \rho^+ \rho^-$ or 
$\to K^* \bar K^*$, etc. Suppose that the magnetic form factor remains positive. (Jlab data suggest that for the proton.) 
If in the spacelike electron-hadron scattering region the electric form factor crosses zero and then approaches zero from below 
with increasing energy, then the relative phase between electric and magnetic form factor is $\pi$. 
But at large timelike momenta, the relative phase between electric and magnetic form factor must then also approach $\pi$ 
with increasing energies, because of the analytic properties of form factors emerging from \eqref{eq:disp1} and the 
high-energy behavior dictated by QCD.
It is a fascinating aspect that timelike data together with rather general mathematical properties of form factors 
can be used to learn something about the spacelike region. This connection has also been highlighted in \cite{Baldini:2005xx}
for the proton. 

Since one has to explore very large energies (timelike momenta) to observe the asymptotic trend of the energy dependence of 
such a relative phase, Belle II \cite{Belle-II:2018jsg} appears to be the appropriate experiment to measure this relative phase. 
The method to extract such a phase is used in \cite{Ablikim:2019vaj} for the $\Lambda$ baryon using data from BESIII. 
Here we want to see if the concept and also the method using angular distributions can easily be extended to vector mesons. 

For spin-1/2 baryons like proton and $\Lambda$ there are two form factors, the magnetic (helicity flip) and 
electric (non-flip) form factors. 
For the vector-meson case one has to study 
three instead of two form factors, related to electric charge, magnetic dipole, and electric quadrupole \cite{Kim:1973ee}. 
We will show that one can determine 
in the timelike region all relative phases between form factors by studying the angular distributions of the decay products of 
the vector mesons. 
Also for vector mesons, a relative phase in the asymptotic timelike region can be related to the relative phase in the 
asymptotic spacelike region. We can obtain information whether specific form factors have a zero crossing in the spacelike 
region. 

To set the stage, we will review in Section \ref{sec:proton} the electromagnetic form factors of the proton and their 
asymptotic behavior. In Section \ref{sec:FFvm} we will define the corresponding form factors for vector mesons. 
We will provide explicit formulae for angular distributions of the decay products from vector mesons produced in 
electron-positron collisions. To provide a comprehensive work we will present formulae that apply to the main decay channels 
of $\rho$, $K^*$, $D^*$, $D^*_s$, $B^*$, and $B^*_s$ (Section \ref{sec:angl-distr}). 
We will summarize our findings in Section \ref{sec:summary}. 
We hope that the present work will convince the experimental colleagues that it is interesting to analyze the corresponding 
data and that the 
explicit formulae for the angular distributions will turn out to be helpful for this endeavor.

\section{Digression to proton form factors}
\label{sec:proton}

Since we have more information about the proton, it is illustrative to discuss this case first and see what one can learn 
from it. The Dirac ($F^p_1$) and Pauli ($F^p_2$) form factors of the proton 
are defined by (see, e.g., \cite{Kubis:2000aa} and references therein)
\begin{eqnarray}
  \langle p(k',\lambda') \vert \, j^\mu \, \vert p(k,\lambda) \rangle 
  = e \, \bar u(k',\lambda') \, \left( \gamma^\mu \, F^p_1(q^2) + \frac{i \sigma^{\mu\nu} \, q_\nu}{2 m_p} \, F^p_2(q^2) 
  \right) \, u(k,\lambda)
  \label{eq:defFF-proton}
\end{eqnarray}
with the electromagnetic current $j^\mu$, the momentum $q=k'-k$ of the virtual photon and the helicities 
$\lambda$ and $\lambda'$. 
These form factors are free from kinematic constraints, in contrast to the electric and magnetic Sachs 
form factors defined by
\begin{eqnarray}
  G^p_E(q^2) & := & F^p_1(q^2) + \frac{q^2}{4 m_p^2} \, F^p_2(q^2) \,, 
  \nonumber \\ 
  G^p_M(q^2) & := & F^p_1(q^2) + F^p_2(q^2)  \,.
  \label{eq:defFFEM-proton}
\end{eqnarray}
Obviously, the latter satisfy the kinematic constraint 
\begin{eqnarray}
  G^p_E(4m_p^2) = G^p_M(4m_p^2)  \,.
  \label{eq:kinemat-constr-proton}
\end{eqnarray}
This kinematic condition does not relate to the reaction $\gamma^* \, p \to p$ but rather to $\gamma^* \to p \, \bar p$.
Indeed, one can also write down the form factors in the timelike region 
\begin{eqnarray}
  \langle \bar p \, p \vert \, j^\mu \, \vert 0 \rangle 
  = e \, \bar u(k,\lambda) \, \left( \gamma^\mu \, F^p_1(q^2) + \frac{i \sigma^{\mu\nu} \, q_\nu}{2 m_N} \, F^p_2(q^2) 
  \right) \, v(k',\lambda')  \,.
  \label{eq:defFF-proton-timelike}
\end{eqnarray}
The photon momentum is now given by the sum of the momenta of the two baryons, $q=k+k'$.

There is also a physical interpretation of \eqref{eq:kinemat-constr-proton}. In the center-of-mass frame of the reaction 
$\gamma^* \to p \, \bar p$, the photon spin orientation measured in the direction of the outgoing proton is given 
by $\lambda_\gamma = \lambda - \lambda'$. A helicity-flip situation, $\lambda_\gamma = \pm 1$, probes $G^p_M$ while a non-flip 
situation, $\lambda_\gamma = 0$, probes $G^p_E$. Thus, $G^p_E$ and $G^p_M$ constitute helicity amplitudes, which are in turn 
related to partial waves \cite{Jacob:1959at}. For the reaction $\gamma^* \to p \, \bar p$, the proton and antiproton can have 
orbital angular momentum $L=0$ (s wave) or $L=2$ (d wave). The helicity amplitudes are superpositions of the s and d wave. 
However, at the production threshold, $q^2 = 4m_p^2$, the d wave is kinematically suppressed. Thus the helicity amplitudes must be
related at this particular point, because both are fully determined by the s wave. This kinematic constraint is nothing but 
\eqref{eq:kinemat-constr-proton}. 

Because of time reversal symmetry, the form factors must be real in the spacelike region of electron-proton scattering: 
$F^p_i(q^2) \in \mathds{R}$ for $q^2 < 0$. But their relative sign could be positive or negative. 
In the timelike region, however, the form factors are complex. Relative phases between 
$F^p_1$ and $F^p_2$ emerge. In the spacelike region, these phases can only be $0$ or $\pi$. 
These properties carry over to the electric and magnetic Sachs form factors introduced in \eqref{eq:defFFEM-proton}.

In addition, one knows the values of the form factors at the photon point:
\begin{eqnarray}
  \label{eq:charge-magnetmom-proton}
  G^p_E(0) = 1 \,, \quad G^p_M(0) = 1 + \kappa \qquad \Leftrightarrow \qquad 
  F^p_1(0) = 1 \,, \quad F^p_2(0) = \kappa 
\end{eqnarray}
with the anomalous magnetic moment $\kappa \approx 1.79$ of the proton \cite{pdg}. 
Thus at the beginning of the spacelike region, there is no relative phase between 
electric and magnetic form factor. If the electric form factor has one zero crossing in the spacelike region and the 
magnetic form factor has none, 
then there is a relative phase of $\pi$ between these two form factors at large negative $q^2$. What does this imply for the 
relative phase in the timelike region? We will show that it must reach $\pi$ at large energies. 

But let us first show that a zero crossing for $G^p_E$ is quite natural. 
At large spacelike $q^2 =: -Q^2$, i.e.\ for $Q^2 \to +\infty$, the constraint-free form factors show the following 
asymptotic behavior \cite{Belitsky:2002kj}: 
\begin{eqnarray}
  \label{eq:scaling-w-logs}
  F^p_1(-Q^2) \sim \frac1{Q^4} \left[\alpha_s(Q)\right]^{2+32/(9\beta)} \,, \qquad
  F^p_2(-Q^2) \sim \frac1{Q^6} \left[\alpha_s(Q)\right]^{2+8/(3\beta)} \log^2(Q^2/\Lambda^2) 
\end{eqnarray}
with the QCD running coupling $\alpha_s$ and leading-order expression of the beta function, $\beta = 11-2 n_f/3$. 
In the massless limit, the helicity-flip part $F^p_2$ involves quark orbital angular momentum components to adjust for the 
helicity flip.
These contributions are infrared sensitive and lead to the $\log^2$ term. The infrared enhancement is cut off by the hadronic 
scale $\Lambda$. 

At $q^2 = 0$ the respective signs of the constraint-free form factors are both positive; see \eqref{eq:charge-magnetmom-proton}. 
Suppose that the Dirac and Pauli form factors are as simple as possible in the 
spacelike region, i.e.\ they only drop. Asymptotically they behave like \eqref{eq:scaling-w-logs}. If both Dirac and Pauli 
form factor are positive in the whole spacelike region, then $G^p_M$ as given in \eqref{eq:defFFEM-proton} 
remains also always positive in the spacelike region. However, the ratio $G^p_E/G^p_M$ is positive at $Q^2 = 0$ but becomes 
negative at large $Q^2$ because of the $\log^2$ term of the Pauli form factor:
\begin{eqnarray}
  \frac{G^p_E(-Q^2)}{G^p_M(-Q^2)} &=& \frac{F^p_1(-Q^2) - \frac{Q^2}{4 m_p^2} \, F^p_2(-Q^2)}{F^p_1(-Q^2) + F^p_2(-Q^2)}
  = \frac{1 - \frac{Q^2}{4 m_p^2} \, \frac{F^p_2(-Q^2)}{F^p_1(-Q^2)}}{1 + \frac{F^p_2(-Q^2)}{F^p_1(-Q^2)}} \nonumber \\
  &\to & - \frac{1}{4 m_p^2} \, \left[\alpha_s(Q)\right]^{8/(9\beta)} \log^2(Q^2/\Lambda^2) \to -\infty 
  \label{eq:ratio-proton-asy}
\end{eqnarray}
where we have dropped positive multiplicative constants. 
Note that it is only logarithms that make the product $Q^2 \cdot F^p_2$ grow faster than $F^p_1$. 
Thus the zero crossing might set in rather late, i.e.\ at rather large $Q^2$. 

This line of reasoning is not a proof that there is a zero crossing in the electric form factor (and non in the magnetic). 
But it is quite plausible. 
And, in principle, there could be several zero crossings instead of just one, but we regard this possibility
as much less plausible. 

Suppose that \eqref{eq:ratio-proton-asy}, a statement for the asymptotic spacelike region, is true. What does this mean for 
the relative phase between electric and magnetic form factor in the {\em timelike} region? For this purpose one can make use of 
the Phragm\'en-Lindel\"of 
method\footnote{As a fun fact we note that the affiliation of the author of this paper coincides with the alma mater of 
  L.E.\ Phragm\'en.}; see, e.g., \cite{Logunov:1963tk}. The asymptotic behavior \eqref{eq:scaling-w-logs} and the 
analytic property \eqref{eq:disp1} of form factors imply
\begin{eqnarray}
  \lim\limits_{q^2 \to \infty} \frac{(q^2)^2 \, F^p_1(q^2)}{(-q^2)^2 \, F^p_1(-q^2)} = 1  \,, \qquad
  \lim\limits_{q^2 \to \infty} \frac{(q^2)^3 \, F^p_2(q^2)}{(-q^2)^3 \, F^p_2(-q^2)} = 1  \,.
  \label{eq:ratio-space-time}
\end{eqnarray}
Note that strictly speaking the form factors have a cut along the positive real axis. By our relation \eqref{eq:disp1} we provide
a well-defined meaning for the form factors on the real axis, i.e.\ we take the value above the cut.

If we write 
\begin{eqnarray}
  \frac{G^p_E(q^2)}{G^p_M(q^2)} 
  = \left\vert \frac{G^p_E(q^2)}{G^p_M(q^2)} \right\vert \, \exp\left[i\left(\Phi_E(q^2)-\Phi_M(q^2)\right)\right]  \,,
  \label{eq:ratio-def-phase-proton}
\end{eqnarray}
i.e.\footnote{We define the phase ``arg'' of a complex number in the range $[0,2\pi)$.}
\begin{eqnarray}
  \Delta\Phi(q^2) := \Phi_E(q^2)-\Phi_M(q^2) = {\rm arg}\left(G^p_E(q^2)/G^p_M(q^2) \right) \,,
  \label{eq:defdelta-Phi-proton}
\end{eqnarray}
we obtain 
\begin{eqnarray}
  \lim\limits_{q^2 \to \infty} \Delta\Phi(q^2) = \lim\limits_{Q^2 \to \infty} \Delta\Phi(-Q^2) = {\rm arg}(-1) = \pi \,.
  \label{eq:asy-delta-Phi-proton}
\end{eqnarray}
Thus, starting from a vanishing relative phase at the production threshold, see \eqref{eq:kinemat-constr-proton}, 
the complex form factors
develop as a function of $q^2$ in such a way that for asymptotically large values the relative phase becomes $\pi$. 
We stress again that this happens under the assumption that \eqref{eq:ratio-proton-asy} is true. 

Albeit constituting an interesting result, it is difficult in practice to measure this relative phase in the reaction 
$e^+ e^- \to p \bar p$, because one needs to measure the helicities of proton and antiproton, 
see, e.g.\ the corresponding discussions in \cite{Baldini:2005xx,Perotti:2018wxm}. Thus one would rather search directly for 
the zero crossing in the spacelike data of electron-proton scattering by going to larger values of $Q^2$. 
However, the situation is somewhat reversed for the case of vector mesons. Here it is practically impossible to study the 
scattering of vector mesons on electrons. But it is possible to study the timelike production region of vector mesons.
In contrast to the proton case, it is possible to extract relative phases between the form factors from the angular decay pattern
of vector mesons. From the asymptotic behavior in the timelike region one can conclude back on a possible zero crossing in the 
spacelike region. In the following, we will provide the formalism to extract this information.

\section{Form factors of vector mesons}
\label{sec:FFvm}

In principle, the definitions and relations provided in this section apply to all mesons with spin 1. 
We will use the strange $K^*$ vector mesons as a concrete example. 

With the later focus on timelike processes, we define constraint-free form factors 
via \cite{Kim:1973ee,Brodsky:1992px,deMelo:2016lwr,Kumar:2019eck}
\begin{eqnarray}
  \label{eq:defVFF}
  \langle K^*(p_1,\lambda_1)  \, \bar K^*(p_2,\lambda_2)  \vert j^\mu \vert 0 \rangle 
  = -e \, \varepsilon^*_\alpha(p_1,\lambda_1) \, \varepsilon^*_\beta(p_2,\lambda_2) \, \Gamma^{\mu\alpha\beta}(q,r)
\end{eqnarray}
with
\begin{eqnarray}
  \Gamma^{\mu\alpha\beta}(q,r)
  & := & F_1(q^2) \, r^\mu g^{\alpha\beta} + F_2(q^2) \left(q^\alpha g^{\mu\beta}-q^\beta g^{\mu\alpha} \right)
  + F_3(q^2) \, \frac{1}{2 m_{K^*}^2} \, r^\mu q^\alpha q^\beta 
  \label{eq:defVFFfinal}
\end{eqnarray}
and $q:=p_1+p_2$, $r:=p_1-p_2$. The helicities of the vector mesons are denoted by $\lambda_{1/2}$. Their mass by $m_{K^*}^2$. 

Before moving on, we would like to discuss a subtlety that we will tentatively ignore in the 
rest of the present work. Strictly speaking, equation \eqref{eq:defVFF} uses implicitly that there are asymptotic 
states $\vert K^*(p,\lambda) \rangle$. This is, however, not quite true, because the vector mesons are unstable resonances. 
In fact, we 
will use in Section \ref{sec:angl-distr} the property that vector mesons decay. The reaction amplitudes where resonances 
appear as intermediate states have poles on one of the Riemann sheets that extend the complex plane in the presence of cuts; 
see, e.g.\ \cite{Belle-II:2018jsg,Hoferichter:2017ftn,Hilt:2017iup}. In this spirit, \eqref{eq:defVFF} can be 
regarded as a reasonable starting point for vector-meson momenta whose square is close to the pole position. In practice, 
the decay widths of most ground-state vector mesons are relatively small (maybe the $\rho$ meson might be regarded as an 
exception). Thus, we decided to ignore this subtlety in the present work. From the experimental point of view, the use of 
asymptotic states in \eqref{eq:defVFF} corresponds to counting ``all'' vector mesons, i.e.\ to integrate over the spectral 
distribution. We will come back to this point in Section \ref{sec:angl-distr}. 

As already spelled out, the form factors of vector mesons can be measured in the timelike ($q^2 > 0$) 
production region in the reaction $e^+ e^- \to K^* \bar K^*$. In the timelike region, the form factors are complex. The imaginary
part is caused by rescattering processes. We will see in the next section that relative phases between form factors can be 
measured. 
In practice, the spacelike ($q^2 <0$) region of elastic scattering $e^- \, K^* \to e^- \, K^* $ cannot be addressed 
experimentally. But the form factors can be defined or analytically extended also to the spacelike region. Here the 
form factors must be real because of time reversal symmetry. Thus, the relative phases can be either $0$ or $\pi$ in the 
spacelike region. 

The decomposition \eqref{eq:defVFF}, \eqref{eq:defVFFfinal} corresponds to \eqref{eq:defFF-proton}. For the timelike region, 
it is convenient to define in the center-of-mass frame the following helicity amplitudes: 
\begin{eqnarray}
  H_{11}(q^2) &:=& F_1(q^2) \,, \qquad H_{10}(q^2) := F_2(q^2) \,, \nonumber \\
  H_{00}(q^2) &:=& \frac{1}{2m_{K^*}^2} 
  \left( (2m_{K^*}^2-q^2) \, F_1(q^2) + q^2 \, F_2(q^2) + q^2 \left(1-\frac{q^2}{4m_{K^*}^2} \right) F_3(q^2)  \right)  \,.
  \label{eq:defhelampH}
\end{eqnarray}
They satisfy the kinematic constraints 
\begin{eqnarray}
  \label{eq:kinconstr111}
  H_{00}(0) = H_{11}(0) \qquad \mbox{and} \qquad H_{00}(4 m_{K^*}^2) = -H_{11}(4 m_{K^*}^2) + 2 H_{10}(4 m_{K^*}^2) \,.
\end{eqnarray}
The labels in $H_{\lambda_1 \, \lambda_2}$ denote the helicities of the vector mesons, cf.\ \eqref{eq:defVFF}. Because of parity 
symmetry the helicity amplitudes $H_{\lambda_1 \, \lambda_2}$ and $H_{-\lambda_1,-\lambda_2}$ are related. 

The differential cross section for $e^+ e^- \to K^* \bar K^*$ is proportional to the following combination:
\begin{eqnarray}
  \frac{d\sigma}{d\cos\theta} &\propto & 
  (s-4m_{K^*}^2) \left( 4m_{K^*}^2 \, (\vert H_{00} \vert^2 + 2 \vert H_{11}\vert^2 ) \sin^2\theta 
    + 2 s \, \vert H_{10} \vert^2 \, (1+\cos^2\theta)  \right)
  \label{eq:propcross}
\end{eqnarray}
with $s=q^2$ and the scattering angle $\theta$ defined in the center-of-mass system. We will provide more differential 
expressions in the next section, accounting for the decays of the vector mesons. 
Here, \eqref{eq:propcross} serves only to illustrate the role of the helicity amplitudes. In the rest frame of the virtual 
photon and for a spin quantization axis along the flight direction of $K^*$, the spin orientation of the virtual photon is 
given by $\lambda_1-\lambda_2$. There is one helicity-flip amplitude $H_{10}$ and two non-flip amplitudes 
$H_{00}$ and $H_{11}$. 

There is a set of other form factors which is more commonly used. 
In line with \cite{Kim:1973ee,Xu:2019ilh} we define the following multipole form factors: 
\begin{eqnarray}
  G_C(q^2) &:=& \frac23 H_{11}(q^2) + \frac13 H_{00}(q^2) \,, \qquad G_M(q^2):= H_{10}(q^2) \,, \nonumber \\[1em] 
  \frac{q^2}{2 m_{K^*}^2} \, G_Q(q^2) &:=& H_{11}(q^2)-H_{00}(q^2) \,.
  \label{eq:defEMQ}
\end{eqnarray}
The electric charge is then $e G_C(0)$, the magnetic moment $\mu = e G_M(0)/(2 m_{K^*})$ and the 
electric quadrupole moment $Q=e G_Q(0)/m_{K^*}^2$. Concerning the multipole 
moments we note in passing that for a positive charge, $G_C(0)=+1$, one interesting aspect is whether the magnetic 
moment of a vector meson is close to the Belinfante conjecture, $G_M(0) = +1$, or close to the value for an elementary gauge 
boson, $G_M(0) = +2$, or completely different from both (see, e.g., the discussion in \cite{Holstein:2006wi}). 

In terms of the constraint-free form factors, the charge, magnetic and quadrupole form factors read 
\begin{eqnarray}
  G_C(q^2) &=& F_1(q^2) - \frac23 \frac{q^2}{4m_{K^*}^2} G_Q(q^2)  \,, \qquad G_M(q^2)=F_2(q^2) \,,  \nonumber \\
  G_Q(q^2) &=& F_1(q^2) - F_2(q^2) + \left(\frac{q^2}{4m_{K^*}^2}-1 \right) F_3(q^2)  \,. 
  \label{eq:F123GEMQ}
\end{eqnarray}
These multipole form factors satisfy the kinematic constraint
\begin{eqnarray}
  \label{eq:constrmultipoles}
  G_Q(4m_{K^*}^2) = 3 \left( G_C(4m_{K^*}^2)-G_M(4m_{K^*}^2) \right) \,.
\end{eqnarray}

It is a matter of taste to regard either the multipole form factors $G_{C/M/Q}$ or the helicity amplitudes 
$H_{\lambda_1 \, \lambda_2}$ as the 
vector-meson analogs of the Sachs form factors \eqref{eq:defFFEM-proton}. 
But we feel obliged to provide some more reasoning for the choice of the multipole form factors $G_{C/M/Q}$:   
\begin{itemize}
\item Resembling the spin-1/2 case (in the timelike region \cite{Leupold:2017ngs}), 
  the electric form factors (charge and quadrupole) are obtained from the non-flip amplitudes,
  the magnetic form factor coincides with the helicity-flip amplitude.
\item The electric (charge) form factor $G_C$ is defined as a spin average of the non-flip helicity amplitudes. 
\item The quadrupole form factor $G_Q$ emerges only if there is a difference between non-flip helicity amplitudes. 
  If there was no difference, the remaining kinematic constraint in \eqref{eq:kinconstr111} would reduce to 
  $G_C(4 m_{K^*}^2) = G_M(4 m_{K^*}^2)$, which resembles the relation \eqref{eq:kinemat-constr-proton} known from the spin-1/2 
  case where there are only two instead of three form factors.
\item The combination 
  \begin{eqnarray}
    \left\vert G_C \right\vert^2 + \frac29 \left\vert \frac{q^2}{2 m_{K^*}^2} \, G_Q \right\vert^2 &=& 
    \frac19 \left\vert 2 H_{11} + H_{00} \right\vert^2 + \frac29 \left\vert H_{11} - H_{00} \right\vert^2 \nonumber \\
    &=& \frac23 \left\vert H_{11} \right\vert^2 + \frac13 \left\vert H_{00} \right\vert^2
    \label{eq:longcalcEQ}
  \end{eqnarray}
  is proportional to the combination that appears in \eqref{eq:propcross} together with $\sin^2 \theta$. 
\end{itemize}
The last aspect tells that the differential cross section \eqref{eq:propcross} shows neither interferences between 
the helicity amplitudes nor between the multipole form factors. In this sense it does not matter much to use 
either the magnetic form factor on the one hand and the 
electric form factors on the other or the helicity-flip amplitude on the one hand and the 
non-flip amplitudes on the other. 
However, we will see in the next section that there is a refinement to this point of view 
when one looks at the decay pattern of the vector mesons. 

We finish this section by discussing the high-energy behavior of the electromagnetic form factors of vector mesons. 
We will be much more schematic than for the proton case. In particular, we will not address in any detail the logarithmic 
corrections to the high-energy scaling based on quark-counting rules. 
Translating \cite{Vainshtein:1977db,Lepage:1980fj,Carlson:1985mm} to our case 
at hand yields 
\begin{eqnarray}
  F_2(-Q^2) = H_{10}(-Q^2) = G_M(-Q^2) \sim  \frac1{Q^4}  \,, \nonumber \\[0.5em]
  F_1(-Q^2) \sim \frac1{Q^4} \,, \qquad  F_3(-Q^2) \sim \frac1{Q^6}  \,,  \nonumber \\[0.5em]
  H_{00}(-Q^2) \sim \frac1{Q^2} \,, \qquad  H_{11}(-Q^2) \sim \frac1{Q^4}  \,,  \nonumber \\[0.5em]
  G_C(-Q^2) \sim \frac1{Q^2} \,, \qquad  G_Q(-Q^2) \sim \frac1{Q^4}  
  \label{eq:scaling-of-all}
\end{eqnarray}
for large $Q^2 = -q^2$. In the spacelike region, the helicity amplitude $H_{00}$ relates to the case where no helicity flip 
of the constituent quarks is required. For $H_{10}$ one needs one and for $H_{11}$ one needs two helicity flips of the 
quarks, caused by mass effects or by orbital angular momentum. 

Somewhat more information can be obtained by the use of light-cone helicity amplitudes \cite{Brodsky:1992px}. For large $Q$, 
one can establish the ratios 
\begin{eqnarray}
  \label{eq:BH-ratio1}
  G_C(-Q^2) \; : \; G_M(-Q^2) \; : \; G_Q(-Q^2) = 
  \left(1-\frac23 \eta \right) \; : \; 2 \left(1+\frac{2\eta-1}{\sqrt{2\eta}} \, {\cal R} \right) \; : \; (-1)
\end{eqnarray}
with $\eta := Q^2/(4m^2)$ and $m$ denoting the vector meson mass. 
Here ${\cal R}$ denotes the ratio of light-cone helicity amplitudes that require one or no quark 
helicity flip, respectively. In the nomenclature of \cite{Brodsky:1992px}: ${\cal R}=G^+_{+0}/G^+_{00}$. 
This ratio is expected to be ${\cal R} \sim \Lambda_{\rm QCD}/Q$ (ignoring again logarithmic corrections). 
If one {\it assumes} furthermore the complete dominance of the quark non-flip amplitude, 
\begin{eqnarray}
  \label{eq:assumpt-r}
  \left\vert \frac{2\eta-1}{\sqrt{2\eta}} {\cal R} \right\vert \ll 1  \,,
\end{eqnarray}
one finds the universal ratios \cite{Brodsky:1992px}
\begin{eqnarray}
  \label{eq:BH-ratio2}
  G_C(-Q^2) \; : \; G_M(-Q^2) \; : \; G_Q(-Q^2) = 
  \left(1-\frac23 \eta \right) \; : \; 2  \; : \; (-1)   \,.
\end{eqnarray}

We stress again that logarithmic corrections have been neglected in \eqref{eq:scaling-of-all}. But these relations are 
sufficient to apply the Phragm\'en-Lindel\"of method/principle and find 
\begin{eqnarray}
  && \lim\limits_{q^2 \to \infty} \frac{(q^2)^2 \, G_M(q^2)}{(-q^2)^2 \, G_M(-q^2)} = 1  \,, \nonumber \\[0.5em]
  && \lim\limits_{q^2 \to \infty} \frac{(q^2)^2 \, F_1(q^2)}{(-q^2)^2 \, F_1(-q^2)} = 1  \,, \qquad 
  \lim\limits_{q^2 \to \infty} \frac{(q^2)^3 \, F_3(q^2)}{(-q^2)^3 \, F_3(-q^2)} = 1  \,, \nonumber \\[0.5em]
  && \lim\limits_{q^2 \to \infty} \frac{q^2 \, H_{00}(q^2)}{(-q^2) \, H_{00}(-q^2)} = 1  \,, \qquad
  \lim\limits_{q^2 \to \infty} \frac{(q^2)^2 \, H_{11}(q^2)}{(-q^2)^2 \, H_{11}(-q^2)} = 1  \,,  \nonumber \\[0.5em]
  && \lim\limits_{q^2 \to \infty} \frac{q^2 \, G_C(q^2)}{(-q^2) \, G_C(-q^2)} = 1  \,, \qquad 
  \lim\limits_{q^2 \to \infty} \frac{(q^2)^2 \, G_Q(q^2)}{(-q^2)^2 \, G_Q(-q^2)} = 1  \,.
  \label{eq:ratio-space-time-vec}
\end{eqnarray}
These relations can be used to deduce how ratios of form factors, their moduli and phases, are related at asymptotic 
spacelike and asymptotic timelike momenta. For instance, the helicity amplitudes $H_{11}$ and $H_{10}$ have the same 
asymptotic behavior (up to logarithmic corrections which do not matter for the Phragm\'en-Lindel\"of 
method \cite{Logunov:1963tk}). Thus one obtains
\begin{eqnarray}
  \lim\limits_{Q^2 \to \infty} {\rm arg}\left(H_{11}(-Q^2)/H_{10}(-Q^2) \right) 
  = \lim\limits_{q^2 \to \infty} {\rm arg}\left(H_{11}(q^2)/H_{10}(q^2) \right) \,.
  \label{eq:ex-rat11}  
\end{eqnarray}

The helicity amplitude $H_{00}$ drops slower than the other helicity amplitudes. This brings in an extra sign! 
To perform the steps in some detail, e.g.\ for the ratio of $H_{00}$ and $H_{10}$:
\begin{eqnarray}
  \lim\limits_{q^2 \to \infty} \frac{H_{00}(q^2)}{-H_{00}(-q^2)} \frac{H_{10}(-q^2)}{H_{10}(q^2)} = 1 
  \label{eq:ex-rat22}  
\end{eqnarray}
and therefore
\begin{eqnarray}
  \lim\limits_{q^2 \to \infty} {\rm arg}\left(H_{00}(q^2)/H_{10}(q^2) \right) &=& 
  \lim\limits_{q^2 \to \infty} {\rm arg}\left(-H_{00}(-q^2)/H_{10}(-q^2) \right) \nonumber \\
  &=& \pi - \lim\limits_{q^2 \to \infty} {\rm arg}\left(H_{00}(-q^2)/H_{10}(-q^2) \right) \,.
  \label{eq:ex-rat22b}  
\end{eqnarray}

Though we focus in the present work on relative phases, a comment about the absolute values of the helicity amplitudes is 
in order. For the observable differential cross section \eqref{eq:propcross}, the amplitude $H_{00}(s)$ dominates at large 
$s=q^2 >0$ (timelike region). This makes sense because only this amplitude does not require any helicity flip on the quark 
level. However, the contribution from $H_{10}$ to the cross section \eqref{eq:propcross} 
is enhanced by a kinematic factor of $s$. Thus there remains only a relative $1/s$ suppression 
between the $s \vert H_{10} \vert^2$ contribution and the dominant $\vert H_{00} \vert^2$ term. 

Let us provide two further examples, one related to the literature, one fitting to the simpler versions of the formulae that 
we will find in the next section. 
In \cite{Xu:2019ilh}, it has been predicted that $G_C$ has a zero crossing in the spacelike region. 
Suppose that the magnetic form factor does not change sign (resembling the proton case). Then the phase of the ratio $G_C/G_M$ 
flips from 0 to $\pi$ at the zero crossing of $G_C$. But the high-energy limit, $Q^2 \to +\infty$, of this relative phase between
$G_C$ and $G_M$ can be related to the high-energy limit in the timelike region. 
So far, we repeat here qualitatively the considerations for the 
proton case, discussed in Section \ref{sec:proton}. However, we will show in the next section that for vector mesons it is 
much easier than for the proton to access this relative phase in the timelike region. Therefore, theoretical predictions 
for zero crossings in the spacelike region can be tested by measurements in the timelike region. 

Another difference to the 
proton case lies in the fact that charge and magnetic form factor have rather different high-energy behavior, 
cf.\ \eqref{eq:scaling-of-all}, while for the proton electric and magnetic Sachs form factors behave the same except for 
logarithmic corrections, cf.\ \eqref{eq:ratio-proton-asy}. Nonetheless, what applies to both proton and vector mesons is the 
idea to check how many zero crossings one might have, e.g.\ one or none, and to check what happens at the photon point, $Q^2=0$.
The signs of the multipole moments at the photon point and the number of zero crossings in the spacelike region tell about the 
relative phases between the form factors at asymptotically large $Q^2$ in the spacelike region. 

As a second example we anticipate that in the context of equation \eqref{eq:angdistr-qlong} below we will find that the 
phase between quadrupole form factor $G_Q(s)$ and magnetic form factor $G_M(s)$ can be determined as a function of $s$. We have 
\begin{eqnarray}
  \lim\limits_{s \to \infty}{\rm arg}\left(G_Q(s) \, G_M^*(s)\right) 
  = \lim\limits_{s \to \infty} {\rm arg}\left(G_Q(s)/G_M(s)\right) 
  = \lim\limits_{Q^2 \to \infty} {\rm arg}\left(G_Q(-Q^2)/G_M(-Q^2)\right)  \,.  \phantom{mm}
  \label{eq:exampleQM}
\end{eqnarray}
The starting expression can be determined experimentally, as we will see in the next section. The final expression can only 
be $\pi$ or zero, because the form factors are real in the spacelike region. 
Thus using \eqref{eq:exampleQM}, experiment can tell whether it is $\pi$ or zero and test in this way
a theoretical prediction about the relative sign between $G_Q$ and $G_M$ at asymptotically large spacelike momenta. The amazing
aspect is that this does not require any measurements in the spacelike region, which are practically impossible. 
Suppose that a theory predicts one zero crossing 
of $G_Q$, none of $G_M$, and positive values of the 
form factors at $Q^2 =0$, i.e.\ positive multipole moments of the considered vector meson. In this case, this theory predicts 
that the expression in \eqref{eq:exampleQM} is $\pi$ and not zero. This prediction can be tested. Of course, experiment cannot 
tell whether one has one or three or any other odd number of zero crossings. 
But we do not expect that form factors of ground-state 
vector mesons have many wiggles in the spacelike region of $q^2 = -Q^2 < 0$.

In general, we expect that in the future, QCD inspired models or even first-principle QCD calculations will provide 
more and more reliable predictions for the electromagnetic multipole moments of vector mesons of various flavor content 
and also predictions for 
possible zero crossings of form factors in the spacelike region. In general, calculations for the spacelike region 
appear to be easier to be performed as compared to calculations in the timelike region where a proper account of the 
inelasticities caused by other many-body states is crucial. But theory predictions for the spacelike region are sufficient
to predict the asymptotic behavior in the timelike region using the timelike-spacelike connection pointed out here. 
The behavior in the timelike region can be determined by angular distributions as we will see next.

\section{Angular distributions}
\label{sec:angl-distr}

To facilitate experimental analyses as much as possible, we have decided to be very detailed in providing the results, in 
particular when it comes to the definitions of the kinematic variables of the reactions. 
The reader who is only interested in the concepts can jump to the summarizing section. 

In the timelike region, the form factors, including their relative phases, 
can be extracted if angular distributions of the full decay chains are measured. 
All vector mesons have electromagnetic form factors except for those that have the quantum numbers of photons. In other words, 
all open-flavor vector mesons have non-vanishing electromagnetic form factors. If one focuses 
on the experimentally established ground-state vector mesons $\rho^+$, $K^{*+}$, $K^{*0}$, $D^{*0}$, $D^{*+}$, $D^{*+}_s$, 
$B^{*+}$, $B^{*0}$, $B^{*0}_s$, it turns out that the main decay channel is a pseudoscalar meson of same open flavor type plus 
either a pion or a photon (the $D^{*0}$ populates both with large branching fractions). The angular distributions are different 
for the two decay channels because in the rest frame of a decaying vector meson a state with two pseudoscalars couples to 
the vector meson's spin orientation $\lambda =0$ while a state of a pseudoscalar and a photon couples to $\lambda = \pm 1$. 
In the following we provide the angular distributions for both decays. 

Since we will not spell out overall normalizations, but just the angular distributions, it is not necessary to fully specify the 
Lagrangians for the decay processes. It is, however, important to stress that there is no ambiguity because each of the 
discussed decays populates exactly one partial wave. Therefore all possible interaction Lagrangians are onshell equivalent. 
Denoting the vector meson by $V$, the pseudoscalar meson of same open flavor type by $P$, the pion by $\pi$ and the photon 
field strength by $F_{\mu\nu}$, the interaction terms can be formulated as $V^\mu \, \bar P \, \partial_\mu \pi$ and 
$\epsilon_{\alpha\beta\mu\nu} V^\alpha \, \partial^\beta \bar P \, F^{\mu\nu}$. For the concrete calculations we have utilized 
Mathematica \cite{Mathematica} and the package Tracer \cite{Jamin:1991dp}. 
To keep the kinematic details as compact as possible, we 
will formulate the processes in terms of flavored mesons and pions. Replacing the pion by photon labels recovers the 
radiative-decay case. To be explicit for the presentation, we use $K^*$ as a vector meson with a two-meson main decay branch, 
$K\pi$, and we use $B^*$ as a vector meson with a radiative main decay branch, $B\gamma$.

Some angular distributions have been determined by BaBar \cite{BaBar:2008fsh} for the reaction $e^- e^+ \to \rho^+ \rho^-$. 
Unfortunately, there were not enough data to provide the multi-differential distributions that are needed to extract the 
relative phases that we address here. 

We start with the case of $e^- e^+ \to K^* \bar K^*$ with a subsequent decay $K^* \to K \pi$, i.e.\ the directions of 
$\bar K^*$, $K$ and $\pi$ are monitored. The angular distribution satisfies 
\begin{eqnarray}
  \frac{d\sigma}{d\cos\theta \, d\Omega_K} &\propto & 
  \vert H_{00} \vert^2 \, \cos^2\alpha \sin^2\theta 
   \nonumber \\ && {}
  + {\rm Re}\left( (H_{11}-H_{00}) \, H_{10}^* \right) \frac{\sqrt{s}}{4m_{K^*}} \cos\beta \sin(2\alpha) \sin(2\theta) 
  \nonumber \\ && {}
  + \vert H_{10} \vert^2 \, \frac{s}{4m_{K^*}^2} \left(1+\cos^2\alpha \cos^2\theta - \cos^2\beta \sin^2\alpha \sin^2\theta \right)  
  \nonumber \\ && {}
  + \vert H_{11} \vert^2 \, \sin^2\alpha \sin^2\theta   \nonumber \\[1em] 
  &=&   \vert H_{00} \vert^2 \, \cos^2\alpha \sin^2\theta 
   \hspace*{18.6em} \vert \sim 1/s^2 
  \nonumber \\ && {}
  + {\rm Re}\left( G_Q \, G_M^* \right) \frac{s^{3/2}}{8m_{K^*}^3} \cos\beta \sin(2\alpha) \sin(2\theta) 
   \hspace*{8.4em} \vert \sim 1/s^{5/2} 
  \nonumber \\ && {}
  + \vert H_{10} \vert^2 \, \frac{s}{4m_{K^*}^2} \left(1+\cos^2\alpha \cos^2\theta - \cos^2\beta \sin^2\alpha \sin^2\theta \right)  
   \hspace*{2.9em} \vert \sim 1/s^3 
  \nonumber \\ && {}
  + \vert H_{11} \vert^2 \, \sin^2\alpha \sin^2\theta  \,.  
   \hspace*{17em} \vert \sim 1/s^4 \phantom{m}
  \label{eq:angdistr-qlong}
\end{eqnarray}
Thus we find an interference between the magnetic (dipole) and the (electric) quadrupole form factor. We have also provided 
the respective high-energy scaling to the very right of each term. Note that the right-hand side is only proportional to the 
differential cross section. We have not provided an overall coefficient that is also $s$ dependent.

In \eqref{eq:angdistr-qlong}, $\theta$ is the scattering angle between electron and $K^*$ in the center-of-mass frame of the electron-positron collision. 
The angles $\alpha$ and $\beta$ refer to the outgoing $K$ 
(angular sphere $\Omega_K$, thus $d\Omega_K = d(\cos\alpha) \; d\beta)$) 
in the rest frame of its mother particle $K^*$. Note that we look here at the further decay of $K^*$, not of $\bar K^*$. 
The coordinate system 
is defined in the following way: The direction of motion of $\bar K^*$ defines the negative $z$-axis. 
(Note that the $\bar K^*$ moves 
in the same direction in the center-of-mass frame of the electron-positron collision and in the rest frame of $K^*$.) 
The electron and the $\bar K^*$ define the collision plane which is identified with the $x$-$z$ plane. The electron momentum 
has a component in the positive $x$ direction. (Note that this does not change by a boost 
between the center-of-mass frame of the electron-positron collision and the rest frame of $K^*$.)

In the center-of-mass frame we find 
\begin{itemize}
\item electron momentum: $q_1=\frac{\sqrt{s}}2 \, (1,\sin\theta,0,\cos\theta)$;
\item positron momentum: $q_2=\frac{\sqrt{s}}2 \, (1,-\sin\theta,0,-\cos\theta)$;
\item total momentum: $q=\sqrt{s} \, (1,0,0,0)$;
\item $K^*$ momentum: $p=\frac{\sqrt{s}}2 \, (1,0,0,\bar\beta)$; 
\item $\bar K^*$ momentum: $k=\frac{\sqrt{s}}2 \, (1,0,0,-\bar\beta)$
\end{itemize}
with the velocity $\bar\beta = \sqrt{1-4 m_{K^*}^2/s}$. 

In the $K^*$ rest frame we have
\begin{itemize}
\item $K^*$ momentum: $p'= m_{K^*} \, (1,0,0,0)$;
\item $K$ momentum: $p_1'=(E_K,p_{\rm dec} \sin\alpha \cos\beta,p_{\rm dec} \sin\alpha\sin\beta,p_{\rm dec} \cos\alpha)$; 
\item $\pi$ momentum: $p_2'=(E_\pi,-p_{\rm dec} \sin\alpha \cos\beta,-p_{\rm dec} \sin\alpha\sin\beta,-p_{\rm dec} \cos\alpha)$
\end{itemize}
with energies and momentum of the decay products given by $E_K=(m_{K^*}^2+m_K^2-m_\pi^2)/(2 m_{K^*})$, 
$E_\pi=(m_{K^*}^2+m_\pi^2-m_K^2)/(2 m_{K^*})$, $p_{\rm dec}=\lambda^{1/2}(m_{K^*}^2,m_K^2,m_\pi^2)/(2 m_{K^*})$ and the K\"all\'en function
\begin{eqnarray}
  \label{eq:kallenfunc}
  \lambda(a,b,c):=a^2+b^2+c^2-2(ab+bc+ac) \,.
\end{eqnarray}
For completeness we even specify the Lorentz boost: $p^\mu = \Lambda^\mu_{\phantom{\mu}\nu} \, p'^\nu$, 
$p_1^\mu = \Lambda^\mu_{\phantom{\mu}\nu} \, p_1'^\nu$, etc.\ with 
\begin{eqnarray}
  \left(\Lambda^\mu_{\phantom{\mu}\nu} \right) = \bar\gamma \left(
    \begin{array}{cccc}
      1 & 0 & 0 & \bar\beta \\
      0 & 1 & 0 & 0 \\
      0 & 0 & 1 & 0 \\
      \bar\beta & 0 & 0 & 1
    \end{array}
  \right)  
  \label{eq:lorentzboost}
\end{eqnarray}
with $\bar\gamma = (1-\bar\beta^2)^{-1/2} = \sqrt{s}/(2 m_{K^*})$.

Next consider the full reaction of 2 to 4 particles: $e^- e^+ \to K^* \bar K^*$ with subsequent decays $K^* \to K \pi$ and 
$\bar K^* \to \bar K \pi$. This introduces two new angles $\gamma$ (like $\alpha$) and $\delta$ (like $\beta$). It is a nice 
exercise to convince oneself that all the independent kinematic variables of the $2 \to 4$ reaction are properly accounted for
when considering $s=q^2$, the scattering angle $\theta$ and the four angles $\alpha$, $\beta$, $\gamma$, and $\delta$.
We will discuss this briefly at the end of this section.

The angular distribution is given by 
\begin{eqnarray}
&&  \frac{d\sigma}{d\cos\theta \, d\Omega_K \, d\Omega_{\bar K}} \propto 
  \vert H_{00} \vert^2 \, \cos^2\alpha \cos^2\gamma \sin^2\theta 
  \hspace*{12.4em} \vert \sim 1/s^2  \nonumber \\ && {}
  - {\rm Re}(H_{00} \, H_{10}^* ) \frac{\sqrt{s}}{2m_{K^*}} \cos\alpha \cos\gamma \sin(2\theta) 
  \nonumber \\ && {} \phantom{+} {} \times 
  \left( \cos\beta \sin\alpha \cos\gamma + \cos\delta \sin\gamma \cos\alpha  \right)
  \hspace*{14.3em} \vert \sim 1/s^{5/2} \nonumber \\ && {}
  - \frac12 {\rm Re}(H_{11} \, H^*_{00}) \cos(\beta-\delta) \sin(2\alpha) \sin(2\gamma) \sin^2\theta  
  \hspace*{10.6em} \vert \sim 1/s^3  \nonumber \\ && {}
  + \vert H_{10} \vert^2 \, \frac{s}{4m_{K^*}^2} 
  \left[ \left( \cos\beta \sin\alpha \cos\gamma + \cos\delta \sin\gamma \cos\alpha \right)^2 \cos^2\theta 
  \right. \nonumber \\ && \hspace*{8em} \left. 
    + \left( \sin\beta \sin\alpha \cos\gamma + \sin\delta \sin\gamma \cos\alpha \right)^2 \right]  
  \hspace*{6.7em} \vert \sim 1/s^3  \nonumber \\ && {}
  +  {\rm Re}(H_{11} \, H_{10}^*) \frac{\sqrt{s}}{2m_{K^*}} \cos(\beta-\delta) \sin\alpha \sin\gamma \sin(2\theta) 
  \nonumber \\ && {} \phantom{+} {} \times \left( \cos\beta \sin\alpha \cos\gamma + \cos\delta \sin\gamma \cos\alpha  \right)
  \hspace*{14.4em} \vert \sim 1/s^{7/2}  \nonumber \\ && {}
  + \vert H_{11} \vert^2 \, \cos^2(\beta-\delta) \sin^2\alpha \sin^2\gamma \sin^2\theta  \,.
  \hspace*{14.6em} \vert \sim 1/s^4 
  \label{eq:angdistr-qlongvery}
\end{eqnarray}
To the very right we have provided again the large-$s$ scaling of the various terms based on the form-factor scaling 
and the explicitly appearing factors of $s$. The first correction to the dominant behavior provides already an interference 
term. 

As a little cross-check we note that this event distribution is always non-negative as it should be. To see this, one can define 
the combination 
\begin{eqnarray}
  \label{eq:defAcombi}
  A &:=& H_{11} \, \cos(\beta-\delta) \sin\alpha \sin\gamma \sin\theta 
  - H_{00} \, \cos\alpha \cos\gamma \sin\theta \nonumber \\
  && {} + H_{10} \, \frac{\sqrt{s}}{2m_{K^*}} 
  \left( \cos\beta \sin\alpha \cos\gamma + \cos\delta \sin\gamma \cos\alpha  \right) \cos\theta \,.
\end{eqnarray}
The distribution in \eqref{eq:angdistr-qlongvery} can be expressed as the sum of squares:
\begin{eqnarray}
  \label{eq:rewritefullPsPs}
  \vert A \vert^2 + \left\vert H_{10} \, \frac{\sqrt{s}}{2m_{K^*}} 
  \left( \sin\beta \sin\alpha \cos\gamma + \sin\delta \sin\gamma \cos\alpha \right) \right\vert^2    \,.
\end{eqnarray}

If $G_M = H_{10}$ does not have a zero crossing, i.e.\ remains positive in the whole spacelike region, then it makes sense to 
normalize quantities with respect to it. Thus we define 
\begin{eqnarray}
  r_1 := \frac{\vert H_{11} \vert}{\vert H_{10} \vert} \,, \qquad 
  r_0 := \frac{\vert H_{00} \vert}{\vert H_{10} \vert} \,, \nonumber \\
  \frac{H_{11}}{H_{10}} =: r_1 \, e^{i\Phi_1} \,, \qquad   \frac{H_{00}}{H_{10}} =: r_0 \, e^{i\Phi_0} \,.
  \label{eq:normalize-wrt-GM}
\end{eqnarray}
This yields    
\begin{eqnarray}
  && \frac{d\sigma}{d\cos\theta \, d\Omega_K \, d\Omega_{\bar K}} \propto  
  r_0^2 \, \cos^2\alpha \cos^2\gamma \sin^2\theta  
  \hspace*{14.3em} \vert \sim s^2  \nonumber \\ && {}
  -r_0 \cos\Phi_0 \, \frac{\sqrt{s}}{2m_{K^*}} \cos\alpha \cos\gamma \sin(2\theta) 
  \nonumber \\ && {} \phantom{+} {} \times \left( \cos\beta \sin\alpha \cos\gamma + \cos\delta \sin\gamma \cos\alpha  \right)
  \hspace*{14.5em} \vert \sim s^{3/2}  \nonumber \\ && {}
  - \frac12 r_1 r_0 \cos(\Phi_1-\Phi_0) \, \cos(\beta-\delta) \sin(2\alpha) \sin(2\gamma) \sin^2\theta  
  \hspace*{8.5em} \vert \sim s  \nonumber \\ && {}
  + \frac{s}{4m_{K^*}^2} 
  \left[ \left( \cos\beta \sin\alpha \cos\gamma + \cos\delta \sin\gamma \cos\alpha \right)^2 \cos^2\theta 
  \right. \nonumber \\ && \hspace*{8em} \left. 
    + \left( \sin\beta \sin\alpha \cos\gamma + \sin\delta \sin\gamma \cos\alpha \right)^2 \right]  
  \hspace*{6.9em} \vert \sim s  \nonumber \\ && {}
  + r_1 \cos\Phi_1 \, \frac{\sqrt{s}}{2m_{K^*}} \cos(\beta-\delta) \sin\alpha \sin\gamma \sin(2\theta) 
  \nonumber \\ && {} \phantom{+} {} \times \left( \cos\beta \sin\alpha \cos\gamma + \cos\delta \sin\gamma \cos\alpha  \right)
  \hspace*{14.5em} \vert \sim s^{1/2}  \nonumber \\ && {}
  + r_1^2 \, \cos^2(\beta-\delta) \sin^2\alpha \sin^2\gamma \sin^2\theta  \,.
  \hspace*{16.4em} \vert \sim s^0  
  \label{eq:angdistr-qlongvery2}
\end{eqnarray}
Obviously one can determine all four quantities $r_1$, $r_0$, $\Phi_1$, and $\Phi_0$ by a fit to the angular distribution 
\eqref{eq:angdistr-qlongvery2} and one still has one cross-check. (There are six terms and one provides the normalization.) 

The $s$-scaling of the various terms in \eqref{eq:angdistr-qlongvery2} is formally different from the statements made 
in \eqref{eq:angdistr-qlongvery}, because of our rescaling performed in \eqref{eq:normalize-wrt-GM}. We stress again that 
we are interested here in the {\it relative} scaling of the terms, not in the overall $s$-dependence of the total cross section.

For very large values of $s$, the distribution \eqref{eq:angdistr-qlongvery2} is dominated by the first term 
$\sim r_0^2 \sim s^2$. However, due a kinematic enhancement factor, the interference term $\sim r_0 \cos\Phi_0$ is 
only suppressed by $1/\sqrt{s}$ relative to the dominant term. 
Therefore it looks promising to read off the asymptotic trend of the phase $\Phi_0(s)$ before 
the interference term gets drowned in the uncertainties. The same statement applies to the interference term between 
$G_Q$ and $G_M$ in \eqref{eq:angdistr-qlong}. 

Since the phases are supposed to reach either zero or $\pi$, their cosines 
will not become small. What matters are the signs of the interference patterns in \eqref{eq:angdistr-qlongvery2} 
and therefore in practice the correct 
interpretation. Thus we have decided to be rather too explicit than too sloppy in spelling out the kinematic details. 

Let us turn to vector mesons with radiative decays. As a concrete example we use open-bottom mesons. In the previous 
kinematic relations one just needs to change $K$ to $B$ labels and $\pi$ to $\gamma$. 
The angular distribution for $e^- e^+ \to B^* \bar B^*$ with a subsequent decay $B^* \to B \gamma$ (but keeping $\bar B^*$)
is given by 
\begin{eqnarray}
  && \frac{d\sigma}{d\cos\theta \, d\Omega_B} \propto  
  \vert H_{00} \vert^2 \, \sin^2\alpha \sin^2\theta 
  \hspace*{17.5em} 
  \vert \sim 1/s^2   \nonumber \\ && {}
  - {\rm Re}\left( (H_{11}-H_{00}) \, H_{10}^* \right) \frac{\sqrt{s}}{4m_{B^*}} \cos\beta \sin(2\alpha) \sin(2\theta) 
  \hspace*{9.7em} 
  \vert \sim 1/s^{5/2}   \nonumber \\ && {}
  + \vert H_{10} \vert^2 \, \frac{s}{4m_{B^*}^2} \left[1 + \cos^2\beta \sin^2\alpha 
    + (1+ \sin^2\beta \sin^2\alpha) \cos^2\theta  \right]  
  \hspace*{5.7em} 
  \vert \sim 1/s^3   \nonumber \\ && {}
  + \vert H_{11} \vert^2 \, (1+\cos^2\alpha) \sin^2\theta  \,. 
  \hspace*{19.7em} 
  \vert \sim 1/s^4  
  \label{eq:angdistr-qlong-gamma}
\end{eqnarray}
Again, we see that one has access to the relative phase between magnetic and quadrupole form factor. 

Now consider the full reaction of 2 to 4 particles: $e^- e^+ \to B^* \bar B^*$ with subsequent decays $B^* \to B \gamma$ and 
$\bar B^* \to \bar B \gamma$. We normalize again with respect to the magnetic form factor. The angular distribution is given by
\begin{eqnarray}
  && \frac{d\sigma}{d\cos\theta \, d\Omega_B \, d\Omega_{\bar B}} \propto 
  r_0^2 \, \sin^2\alpha \sin^2\gamma \sin^2\theta 
  \hspace*{19.3em} \vert \sim s^2 \nonumber \\[0.5em] && {}
  +r_0 \cos\Phi_0 \, \frac{\sqrt{s}}{2m_{B^*}} \sin\alpha \sin\gamma \sin(2\theta) 
  \left( \cos\beta \cos\alpha \sin\gamma + \cos\delta \cos\gamma \sin\alpha  \right)
  \hspace*{5.8em} \vert \sim s^{3/2} \nonumber \\ && {}
  - \frac12 r_1 r_0 \cos(\Phi_1-\Phi_0) \, \cos(\beta-\delta) \sin(2\alpha) \sin(2\gamma) \sin^2\theta  
  \hspace*{13.1em} \vert \sim s \nonumber \\ && {}
  + \frac{s}{4m_{B^*}^2} 
  \left\{ \left[\left( \cos\beta \cos\alpha \sin\gamma + \cos\delta \cos\gamma \sin\alpha  \right)^2 
      + \sin^2\beta \sin^2\gamma + \sin^2\delta \sin^2\alpha \right] \cos^2\theta 
  \right. \nonumber \\ && \hspace*{4.5em} \left. 
    + \left( \sin\beta \sin\alpha \cos\gamma + \sin\delta \sin\gamma \cos\alpha  \right)^2 
    + \cos^2\beta \sin^2\alpha + \cos^2\delta \sin^2\gamma 
  \right\}  
  \hspace*{2.4em} \vert \sim s \nonumber \\ && {}
  - \frac12 r_1 \cos\Phi_1 \, \frac{\sqrt{s}}{2m_{B^*}} \left[2 \cos(\beta-\delta) \cos\alpha \cos\gamma 
    \left( \cos\beta \cos\alpha \sin\gamma + \cos\delta \cos\gamma \sin\alpha \right)
  \right. \nonumber \\ && {} \hspace*{8.5em} 
  \left. {}+ \sin(\beta-\delta) \left( \sin\beta \sin(2\gamma) - \sin\delta \sin(2\alpha) \right) \right]
  \sin(2\theta) 
  \hspace*{6em} \vert \sim s^{1/2} \nonumber \\ && {}
  + r_1^2 \left[ \cos^2\alpha + \cos^2\gamma 
    + \cos^2(\beta-\delta) \sin^2\alpha \sin^2\gamma \right] \sin^2\theta  \,.   
  \hspace*{12.9em} \vert \sim s^0   \nonumber \\ && {}
  \label{eq:angdistr-qlongvery2-2gammas}
\end{eqnarray}
This angular distribution differs from the two-meson decay case \eqref{eq:angdistr-qlongvery2}, but the qualitative finding 
is the same. For very large values of $s$, the distribution \eqref{eq:angdistr-qlongvery2-2gammas} is dominated by the 
first term $\sim r_0^2 \sim s^2$. The interference term $\sim r_0 \cos\Phi_0$ is 
only suppressed by $1/\sqrt{s}$ relative to the dominant term. 

Again we want to check that the event distribution is always non-negative. To this end we define
\begin{eqnarray}
  \label{eq:defAcombi2}
  \tilde A &:=& H_{00} \, \sin\alpha \sin\gamma \sin\theta 
  - H_{11} \, \cos(\beta-\delta) \cos\alpha \cos\gamma \sin\theta \nonumber \\ && {}
  + H_{10} \, \frac{\sqrt{s}}{2m_{B^*}} 
  \left( \cos\beta \cos\alpha \sin\gamma + \cos\delta \cos\gamma \sin\alpha  \right) \cos\theta  \,,
\end{eqnarray}
\begin{eqnarray}
  \label{eq:defBcombi2}
  B:= -H_{11} \, \sin(\beta-\delta) \cos\gamma \sin\theta + H_{10} \, \frac{\sqrt{s}}{2m_{B^*}} \sin\beta \sin\gamma \cos\theta \,,
\end{eqnarray}
and
\begin{eqnarray}
  \label{eq:deftildeBcombi2}
  \tilde B:= H_{11} \, \sin(\beta-\delta) \cos\alpha \sin\theta 
  + H_{10} \, \frac{\sqrt{s}}{2m_{B^*}} \sin\delta \sin\alpha \cos\theta
\end{eqnarray}
where we have switched back to the helicity amplitudes. The event distribution in \eqref{eq:angdistr-qlongvery2-2gammas} is then 
proportional to the following sum of squares:
\begin{eqnarray}
  && \vert \tilde A \vert^2 + \vert B \vert^2 + \vert \tilde B \vert^2 
  + \left\vert H_{11} \right\vert^2 \cos^2(\beta-\delta) \sin^2\theta  \nonumber \\[0.5em] && {}
  + \left\vert H_{10} \, \frac{\sqrt{s}}{2m_{B^*}} \right\vert^2 
  \left[ \left( \sin\beta \sin\alpha \cos\gamma + \sin\delta \sin\gamma \cos\alpha  \right)^2 
    + \cos^2\beta \sin^2\alpha + \cos^2\delta \sin^2\gamma \right]   \,.  \phantom{m}
  \label{eq:ABtildessq}  
\end{eqnarray}

In all the calculations we have treated the intermediate vector mesons as if they were stable. This corresponds to the 
experimental procedure to integrate over the vector-meson peak. This procedure can be mimicked in the following way. 
Including the width $\Gamma$ of the vector meson in its propagator, 
the spin average of the squared matrix element of the reaction is given by 
\begin{eqnarray}
  \left\langle \vert {\cal M} \vert^2 \right\rangle &=:& \left\langle \vert {\cal M} \vert^2 \right\rangle_{\rm red} \,
  \left\vert \frac1{p^2-m_V^2 + i m_V \Gamma} \right\vert^2 \, 
  \left\vert \frac1{k^2-m_V^2 + i m_V \Gamma} \right\vert^2 \,   \nonumber  \\
  &=& \left\langle \vert {\cal M} \vert^2 \right\rangle_{\rm red} \, \frac1{m_V^2 \Gamma^2} \, 
  \frac{m_V \Gamma}{(p^2-m_V^2)^2 + m_V^2 \Gamma^2} \,   \frac{m_V \Gamma}{(k^2-m_V^2)^2 + m_V^2 \Gamma^2}   \nonumber  \\
  &\approx& \left\langle \vert {\cal M} \vert^2 \right\rangle_{\rm red} \, \frac1{m_V^2 \Gamma^2} \, \pi \delta(p^2-m_V^2) \, \pi 
  \delta(k^2-m_V^2) \,.
  \label{eq:matr-red-amtr}
\end{eqnarray}
Here we have tacitly assumed that the reduced squared matrix element $\left\langle \vert {\cal M} \vert^2 \right\rangle_{\rm red}$ 
does not depend (very much) on the invariant masses $p^2$ and $k^2$ of the vector mesons in the range covered by the 
respective spectral distribution of the invariant mass. The delta functions of \eqref{eq:matr-red-amtr} enter then just 
the phase space integrals that in turn lead to the angular distributions given above. 

As a final cross-check, let us count the number of independent angles: 
A reaction of 2 into 4 particles has $3 \times 6 - 10=8$ independent 
variables.\footnote{The number 10 emerges 
  from energy-momentum conservation and the possibility to change the frame of reference and coordinate system by three boosts 
  and three Euler angles.} 
For given $s=q^2$ and for onshell vector mesons, one gets 5 independent angles. They have been denoted by $\theta$, $\alpha$, 
$\beta$, $\gamma$, and $\delta$. If the decay products of one vector meson are not measured, one has to deal with a reaction 
of 2 into 3 particles. There one has $3 \times 5 - 10=5$ independent variables. This leads to 3 angles, namely $\theta$, 
$\alpha$, and $\beta$. 

\section{Summary}
\label{sec:summary}

We have introduced relative phases between helicity amplitudes in \eqref{eq:normalize-wrt-GM}. 
The $s=q^2$ dependences of these relative phases $\Phi_1$ and $\Phi_0$ are the quantities of interest if one wants to relate 
the asymptotic spacelike and timelike behavior and learn something about zero crossings in the spacelike region. Obviously, 
both relative phases are accessible by analyzing the data with \eqref{eq:angdistr-qlongvery2} or 
\eqref{eq:angdistr-qlongvery2-2gammas}, respectively. There is even some phase information left 
if only one decay distribution is measured, as shown in \eqref{eq:angdistr-qlong} and \eqref{eq:angdistr-qlong-gamma}, 
respectively. In this case, one still has access to 
one particular relative phase, the one between the magnetic and the quadrupole form factor. 
Only for the plain vector-meson production without any differential information about the 
decay products, all phase information is washed out, as can be seen in \eqref{eq:propcross}. 

If one wants to test the prediction of a zero crossing of a particular form factor, we suggest the following procedure: 
On the theory side, it is helpful to identify (if possible) another form factor that remains positive in the spacelike region. 
This defines a relative phase between the two form factors, which is either $\pi$ or zero for spacelike energies beyond the 
zero crossing. With the Phragm\'en-Lindel\"of method/principle, one can translate this information to the case of asymptotic 
timelike momenta. On the experimental side, one can determine the energy dependence of this relative phase in the timelike 
region and test the theoretical prediction. It would be interesting to carry out the experimental program for all 
ground state vector mesons of various open flavor and study in this way the impact of quark mass and quark charge on the 
electromagnetic structure of vector mesons. 

{\bf Acknowledgments:} This work is dedicated to Simon Eidelman. The author will miss 
Simon's incredible physical insights and the smile in his eyes during discussions. 
The interest of the author in the timelike-spacelike connection was sparked by very interesting discussions 
with Simone Pacetti. The author would like to thank Andrzej Kup\'s\'c for general discussions and for his valuable 
comments on the manuscript. This work has been supported by the Swedish Research Council (Vetenskapsr\aa det) 
(grant number 2019-04303).

\bibliography{lit}{}
\bibliographystyle{apsrev4-1}

\end{document}